\documentclass[letterpaper]{article}
\usepackage{aaai25}
\usepackage{times}
\usepackage{helvet}
\usepackage{courier}
\usepackage[hyphens]{url}
\usepackage{graphicx}
\usepackage{natbib}
\usepackage{caption}
\usepackage{lipsum}

\usepackage[linesnumbered,lined,ruled,vlined]{algorithm2e}
\usepackage{algpseudocode}

\usepackage{amsmath,amssymb,amsfonts}
\usepackage{mathtools}
\usepackage{booktabs}
\usepackage{tabularx}
\usepackage{enumitem}
\usepackage{graphicx}
\usepackage{subcaption}
\usepackage[utf8]{inputenc}
\usepackage{multirow}
\usepackage{array}

\title{Task and Perception-aware Distributed Source Coding for Correlated Speech under Bandwidth-constrained Channels}
\author {
    Sagnik Bhattacharya\textsuperscript{\rm 1},
    Muhammad Ahmed Mohsin\textsuperscript{\rm 1},
    Ahsan Bilal\textsuperscript{\rm 2},
    John M. Cioffi\textsuperscript{\rm 1}
}

\affiliations {
\textsuperscript{\rm 1}Stanford University, 450 Jane Stanford Way, Stanford, CA 94305-2004, USA \\
\textsuperscript{\rm 2}University of Oklahoma, 660 Parrington Oval, Norman, OK 73019, USA \\
sagnikb@stanford.edu,
muahmed@stanford.edu,
ahsan.bilal-1@ou.edu,
cioffi@stanford.edu
}

\begin{document}

\maketitle

\begin{abstract}
Emerging wireless AR/VR applications require real-time transmission of correlated high-fidelity speech from multiple resource-constrained devices over unreliable, bandwidth-limited channels. Existing autoencoder-based speech source coding methods fail to address the combination of the following - (1) dynamic bitrate adaptation without retraining the model, (2) leveraging correlations among multiple speech sources, and (3) balancing downstream task loss with realism of reconstructed speech. We propose a neural distributed principal component analysis (NDPCA)-aided distributed source coding algorithm for correlated speech sources transmitting to a central receiver. Our method includes a perception-aware downstream task loss function that balances perceptual realism with task-specific performance. Experiments show significant PSNR improvements under bandwidth constraints over naive autoencoder methods in task-agnostic (19\%) and task-aware settings (52\%). It also approaches the theoretical upper bound, where all correlated sources are sent to a single encoder, especially in low-bandwidth scenarios. Additionally, we present a rate-distortion-perception trade-off curve, enabling adaptive decisions based on application-specific realism needs.
\end{abstract}

%

\section{Introduction}
\label{sec:intro}

Upcoming use cases in wireless augmented reality (AR) and virtual reality (VR)~\cite{wirelessVR}, as well as other immersive applications, often require real-time transmission of high-fidelity speech data from resource-constrained edge devices, such as AR glasses or VR headsets, over inherently unreliable and bandwidth-limited wireless channels~\cite{wirelessAR}. Efficient and adaptive transmission methods, which can handle dynamic channel conditions and limited computational resources, are critical to prevent disruptions that can degrade the immersive experience. Reliable speech source coding for transmission over unreliable dynamic wireless channels is thus the need of the hour~\cite{semanticComms, communicationCompression, jsccImage}.

Current research in deep learning-aided source coding \cite{wirelessDeepSpeech, audiodec, talkingHead, jsccSurvey} has advanced compression strategies for speech data under fixed maximum output bitrates. However, three critical aspects for practical variable bitrate compression in real-world wireless systems remain overlooked: (1) A robust framework for deriving maximum permissible bitrate based on wireless channel characteristics, such as dynamic channel capacity computed from channel state information (CSI), is missing. (2) Existing autoencoder-based methods are designed for fixed bitrates with predetermined output dimensions, requiring retraining for every new dimension, which is impractical in dynamic environments. (3) Scenarios with multiple correlated sources, like speech captured by devices such as VR headsets, AR glasses, and smartphones, are largely ignored. Exploiting correlations among these sources can improve compression efficiency and bandwidth utilization, but most methods treat sources independently. While some works address correlated sources~\cite{hyeji, neuralDistributedSourceCoding}, these techniques have not been applied to speech data.

Recent advancements in source coding have moved beyond traditional methods focused solely on bit reconstruction accuracy \cite{perceptionVideo, rateDistortionPerception}. For AR/VR and next-generation wireless applications, where tasks like speech enhancement, source separation, cloning, and generation are critical, task-aware source coding directly enhances task-specific performance. This approach enables higher compression rates, improved bandwidth efficiency, and superior task performance. For speech source coding, preserving perceptual realism is equally essential \cite{rateDistortionPerception}. Reconstructed speech must sound natural to maintain authenticity, making realism preservation a vital complement to task-aware optimization. Perception-aware source coding has introduced perceptual loss to maintain audio realism, showing promising results. However, the interaction between perceptual loss and task-aware optimization in distributed correlated source coding for speech remains unexplored. Addressing this gap offers an opportunity to design systems that balance task performance and perceptual realism in dynamic, bandwidth-constrained wireless channels.
\begin{figure*}[t!]
    \centering
    \includegraphics[width=1\linewidth]{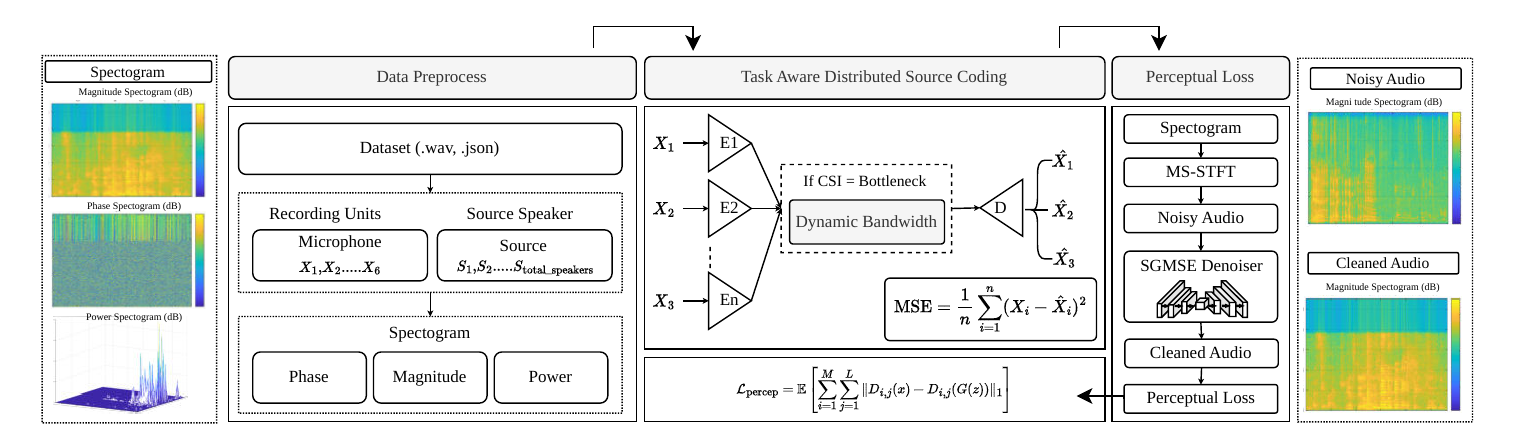}
    \caption{End to end proposed pipeline for distributed downstream speech enhancement using perceptual loss}
    \label{fig:Flow_Chart}
\end{figure*}

In this paper, we propose a neural distributed principal component analysis (NDPCA)-aided source coding algorithm for correlated speech sources transmitting to a central receiver. We also introduce a perception-aware downstream task loss function for training learner encoders and decoders. Experimental evaluations demonstrate:
\begin{itemize}
    \item The proposed NDPCA-aided distributed source coding algorithm achieves 19\% higher average PSNR in task-agnostic settings under bandwidth constraints compared to naive autoencoder-based source coding.
    \item In task and perception-aware settings, the proposed algorithm achieves 52\% higher PSNR compared to the baseline and approaches the performance of the upper bound (all sources combined at one encoder), especially in low-bandwidth scenarios.
    \item The rate-distortion-perception trade-off curve illustrates the algorithm’s ability to balance perception and reconstruction errors, enabling adaptive decision-making under varying realism constraints.
\end{itemize}
These results highlight the algorithm’s effectiveness in optimizing compression, realism, and task performance under bandwidth-constrained scenarios.

\section{Related Work}
\label{sec:related}

\subsection{Wireless CSI based rate adaptation/source coding}
\cite{jsccSurvey} provides a comprehensive survey of wireless CSI-based rate adaptation methods used in current research \cite{rateAdaptiveCoding}. \cite{talkingHead} proposes and implements a neural joint source-channel coding-based talking head semantic transmission system for varying SNR levels, however, the bandwidth is assumed to be largely constant and pre-provided. \cite{vbrAudio} implements reinforcement learning (RL)-based variable bitrate video chunk sizing for efficient transmission, however, assuming the bandwidth to be pre-provided, and not computing it from wireless channel capacity measurements.

\subsection{Speech source coding}
Current research extensively explores single source single receiver source coding for speech data. Works like \cite{wirelessDeepSpeech, audiodec, enhanceCodec} implement efficient source coding under noisy channels, while \cite{perceptionVideo, rateDistortionPerception} analyze rate-distortion-perception trade-offs. However, these approaches rely on fixed output encoder representations followed by entropy coding, which is less efficient than variable output dimensional representations. Designing such variable output representations with existing methods would require creating new encoder models for every output dimension, making it impractical. Additionally, most research targets variational autoencoder-based methods, unlike our proposed algorithm, which employs NDPCA-aided autoencoder mechanisms for fixed representations instead of stochastic latent-based generation.

\subsection{Distributed source coding}
There is a lack of prior work on distributed source coding for speech data under unreliable channel conditions. \cite{hyeji} introduces and implements a distributed source coding algorithm utilizing the correlation between multiple participating sources using NDPCA, and this becomes the basis of our proposed algorithm. We utilize the NDPCA-aware distributed source coding mechanism from \cite{hyeji}, while modifying the distributed autoencoder design and introducing task-perception loss for speech-specific sources. In summary, there is a lack of current research which implements practical downstream task-aware distributed source coding schemes under dynamic bandwidth wireless channels, specifically targeted to speech data, and preserves high perceptual quality at the receiver post decoding.

\section{Wireless Channel-Aware Distributed Source Coding}
\label{sec:distributed}
\subsection{Channel State Information (CSI)-aware Dynamic Bitrate}
Fig.~\ref{fig:floorplan} illustrates a distributed source coding scenario where multiple microphones at different room locations record the same conversational speech. Each source encodes its recorded speech and transmits it to a central receiver for decoding and downstream tasks, such as speech enhancement. We assume perfect wireless CSI availability at the receiver, which is crucial for efficient source coding. Fluctuating channel capacity, inferred from the CSI, determines the total uplink bitrate which has to be optimally allocated across all the sources. The variable dimension encoder output is coded for transmission using entropy coding. From rate-distortion \cite{zamir2008achieving}, we have
\begin{align}
    R(D) \leq \frac{1}{2}\log_2 (\frac{\sigma^2}{D})
\end{align}
where $R(D)$ represents the minimum number of bits per symbol required to maintain a quantization distortion less than or equal to $D$, and $\sigma^2$ is the source variance. 

Now, if the encoder output at source $s$ at time $t$ has dimension $l_{s,t}$, and the outputs are sent once every $T$ second, the bitrate required by that source is given by
\begin{align}
    E_{s,t} = \frac{l_{s,t} \eta R(D)}{T} 
        \leq \frac{l_{s,t} \eta \log_2 (\frac{\sigma^2}{D})}{2T}
\end{align}
where $\eta$ is the number of symbols per floating point output. For the total uplink bit rate constraint to be met, the sum of the bit rates of the individual sources should be less than or equal to the total capacity of the channel at time $t$, measured as $C_t$ at the receiver. Therefore, the sum of the encoder output dimensions of all the sources can be given as
\begin{align}
\label{eq:rate_capacity}
    \sum_{s}l_{s,t} \leq \sum_s l_{\text{Gaussian}, s, t} = \frac{2C_t T}{\eta \log_2 (\frac{\sigma^2}{D})}
\end{align}
We design the encoders for the worst case scenario, that is, where the encoder output follows a Gaussian distribution, and thus keep the number of output dimensions to be time-varying and equal to $l_{\text{Gaussian},s, t }$. Fig.~\ref{fig:bitrate} shows an experimentally obtained graph of channel capacity varying across time in a real-world wireless channel, and the sum of output dimensions of all the source encoders varying accordingly, as in Eq.~\eqref{eq:rate_capacity}. To simply things, henceforth in the paper, by bandwidth, we would refer to the number of output floating point numbers that are being transmitted/received. 

\subsection{NDPCA-aided Distributed System Model}
We design a neural distributed spectral autoencoder with encoders at the transmitting sources and a decoder at the central receiver, as shown in Fig.~\ref{fig:Flow_Chart}. The autoencoder processes the spectrogram’s magnitude and phase, which include frequency bins, channels, batch size, and timestamps. To handle real-time dynamic bandwidth allocation without having to retrain the autoencoder model for each new output dimension requirement, we apply neural distributed principal component analysis (NDPCA) \cite{hyeji} on the latent embeddings from the encoders. The architecture details of the encoder, decoder, and the NDPCA mechanism for efficient, dynamic output dimensions are elaborated in the following paragraphs.

\subsubsection{NDPCA-aided Distributed Encoder.} The designed distributed neural encoder comprises two steps -- an individual multichannel spectral encoder at each of the sources, followed by one distributed PCA encoder.\\
\textbf{(a) Individual Spectral Encoder.}
We use a multichannel spectral autoencoder leveraging a deep residual-based encoder-decoder architecture for feature extraction, representation learning, and data reconstruction. The encoder architecture (\(\mathcal{E}\)) employs a deep residual network with an initial frequency projection followed by convolutional and residual blocks. The input tensor \(\mathbf{x} \in \mathbb{R}^{B \times C \times F \times T}\), representing batch size \(B\), channels \(C\), frequency bins \(F\), and time steps \(T\), is initially reshaped to handle the spectral dimensions effectively.\\
\begin{enumerate}
    \item \textbf{Initial Frequency Projection}: We project the frequency dimension to a lower dimension:
\begin{equation}
    \mathbf{f} = \text{ReLU}(\mathbf{W_2} \cdot \text{ReLU}(\mathbf{W_1} \cdot \mathbf{x})) \in \mathbb{R}^{B \times C \times 128 \times T}
    \end{equation}
    where \(\mathbf{W_1} \in \mathbb{R}^{F \times 256}\) and \(\mathbf{W_2} \in \mathbb{R}^{256 \times 128}\) are learned weights.
    \item \textbf{Residual and Convolutional Layers}: The frequency-projected data is passed through temporal convolutional layers and residual blocks. Each residual block \(\mathcal{R}_i\) introduces non-linearity and stability in feature transformation, defined as:
   \begin{equation}
    \mathcal{R}_i(\mathbf{h}) = \text{ReLU}(\mathcal{L}_2(\text{Conv}(\text{ReLU}(\mathcal{L}_1(\text{Conv}(\mathbf{h}))))))
    \end{equation}
    where \(\mathcal{L}_1\) and \(\mathcal{L}_2\) are layer normalization layers. The output of the convolutional layers and residual blocks is then flattened and transformed to the latent space \(\mathbf{z} \in \mathbb{R}^{B \times Z}\).
\end{enumerate}
\textbf{(b) Distributed PCA encoder} We train a single distributed encoder-decoder model, as described above, which would be used for variable bit rate requirements. We derive the total uplink bandwidth, $B$ derived from channel capacity at the central receiver. The encoder output latent vector $\mathbf{z}_i \in \mathbb{R}^{v_i} ~\forall i \in \{1, 2, \ldots S\}$ at each source, where $S$ is the total number of sources and $v_{i}$ is the output dimension of the $i^{th}$ source's encoder. PCA is applied to each of the source encoder output latent vectors, followed by a distributed selection of the $B$ maximum correlation components from across all the PCA components across all the sources.
\begin{equation}
\mathbf{\hat{z}_i = \mathbf{U_i} \cdot \mathbf{\Sigma_i} \cdot \mathbf{V_i}^T}
\end{equation}
where \(\mathbf{U_i}\), \(\mathbf{\Sigma_i}\), and \(\mathbf{V_i}^T\) are the principal vectors and principal component correlations obtained at the $i^{th}$ source. On selecting the $B$ maximum PCA components from across all the components from all the sources, we get the optimum correlations (the selected singular values), the optimum directions to transform the latent vectors to (the corresponding columns of $U$), as well as the individual allocated optimal bandwidth for each source (number of components chosen from each source). The selected PCA projection vetors from each of the sources are then transmitted over wireless channels to the single receiver decoder, which concatenates the individual incoming vectors to construct the decoder input, $\hat{\mathbf{z}}$.

\subsubsection{Decoder}The Spectral Decoder (\(D\)) reconstructs the original data from the latent representation ($\hat{\mathbf{z}}$), aiming to achieve perceptual fidelity in the spectro-temporal domain. The decoder inverts the transformations applied by the encoder, utilizing the latent vector for reconstructing spectral data.

\begin{enumerate}
    \item \textbf{Latent Projection and Residual Blocks}: The decoder first projects the latent space vector back to the temporal resolution:
    \begin{equation}
    \mathbf{x}_0 = \text{ReLU}(\mathbf{W} \cdot \mathbf{\hat{z}}) \in \mathbb{R}^{B \times 128 \times T}
    \end{equation}
    where \(\mathbf{W} \in \mathbb{R}^{Z \times (128 \times T)}\) is a learned weight. The temporal representations are processed through residual blocks similar to the encoder.

    \item \textbf{Reconstruction with Frequency Projection}: Finally, the data is reshaped and projected back to the frequency dimension:
    \begin{equation}
    \mathbf{x}_{\text{out}} = \text{Conv2D}(\mathbf{W}_{\text{freq}} \cdot \mathbf{x}_0) \in \mathbb{R}^{B \times C \times F \times T}
    \end{equation}
    where \(\mathbf{W}_{\text{freq}}\) maps the temporal features back to the spectral domain, ensuring that the output maintains the original structure of the data.
\end{enumerate}

\section{Task and Perception-aware Loss Function}
\label{sec:perception}
Directly optimizing the end to end NDPCA-aided distributed encoder-decoder model for the required downstream task enables better compression performance than task-agnostic source coding. However, since we are dealing with human speech reconstruction at the central receiver, it is also crucial to preserve realism, that is, the reconstructed speech at the decoder output should sound human speech-like. To that end, we propose a task and perception-aware loss function to train the distributed encoder-decoder pipeline. The loss function comprises three parts - a task-agnostic loss, a downstream task loss, as well as a perceptual loss component. We formulate each of these components in the subsequent subsections, after explaining an initial feature extraction on the original speech data, which is done in the very beginning of the entire pipeline.

\subsubsection{Feature Extraction.}The initial step in this architecture is transforming the speech signal into the time-frequency domain using STFT given as:
\begin{equation}
X(f, \tau) = \sum_{n=-\infty}^{\infty} x(n) \cdot w(n - \tau) e^{-j 2 \pi f n}
\end{equation}
where \(X(f, \tau)\) represents the time-frequency domain representation of the discrete-time signal \(x(n)\). Here, \(w(n - \tau)\) is a window function centered at \(\tau\), enabling localized analysis of \(x(n)\), and \(e^{-j 2 \pi f n}\) is a complex exponential isolating the frequency component \(f\). The STFT captures how the frequency content of \(x(n)\) evolves over time. We use STFTs as source encoder inputs, and recover reconstructed STFTs as decoder outputs. The STFT window length  was 2048 with hop length 512 and window length 2048.
\subsection{Task Agnostic Loss } The task agnostic loss for the distributed autoencoder model is divided into following subparts:
\begin{enumerate}
    \item \textbf{Mean Squared Error (MSE): }The MSE loss calculates the average squared difference between the original ground truth speech data, that is, the STFT matrix of the clean speech, and the reconstructed STFT at the output of the decoder at the central receiver. For ground truth speech STFT $\mathbf{x}_\text{gt}$ and reconstructed decoder output STFT $\mathbf{x}_{\text{dec}}$, the MSE loss $L_{\text{MSE}}$ is defined as:

\begin{equation}
L_{\text{MSE}} = \frac{1}{2} \mathbb{E} \left[ \sum_{i=1}^{B} \sum_{j=1}^{F} \sum_{k=1}^{T} \left( \mathbf{x}_{\text{gt},ijk} - \mathbf{x}_{\text{dec},ijk} \right)^2 \right]
\end{equation}

where $B$ is the batch size, $F$ is the frequency dimension, $T$ is the temporal dimension, and the expectation is over the dataset. The MSE loss is representing the simple task-agnostic reconstruction loss.

\item \textbf{Cosine Similarity Loss: }We want to train the NDPCA-aided distributed encoder-decoder model such that the higher compression performance is achieved utilizing the correlation between the sources. To ensure this, we incorporate a subcomponent in the loss function which would penalize high correlation between the latent embedding vector output by the different source encoders. Cosine similarity compares the similarity between two latent vectors $z_1$ and $z_2$, providing a measure of directional alignment. The Cosine loss $L_{\text{cos}}$ is:
\begin{equation}
L_{\text{cos}} = 1 - \mathbb{E} \left[ \sum_{i,j \in \{1,2,\ldots S\}} \frac{z_i \cdot z_j}{\| z_i \| \| z_j \|} \right]
\end{equation}
where $S$ is the total number of sources, and $z_i$ is the encoder output for the $i^{th}$ source.

\item \textbf{Spectral SNR: }Spectral SNR measures the fidelity of the reconstructed spectral data relative to the original, in decibels (dB). The spectral SNR loss is given by:
\begin{equation}
L_{\text{SNR}} = 10 \log_{10} \left( \frac{\mathbb{E}\left[ (\mathbf{x}_\text{gt} - \mathbf{x}_\text{dec})^2 \right]}{\mathbb{E}\left[ {\mathbf{x}_\text{gt}}^2 \right]} \right)
\end{equation}
\item \textbf{Peak Signal-to-Noise Ratio (PSNR): }PSNR represents the ratio between the maximum possible value and the reconstruction error in dB. The PSNR loss is given by:
\begin{equation}
L_\text{PSNR} = -10 \log_{10} \left( \frac{\mathbf{x}_{\text{max}}^2}{L_{\text{MSE}}} \right)
\end{equation}
where $\mathbf{x}_\text{max}$ is the maximum possible value of the ground truth speech STFT. Each of the aforementioned subparts  contributes uniquely to achieving the balance between accurate spectral reconstruction and regularized latent representations. 
\end{enumerate}

\begin{figure}
    \centering
    \includegraphics[width=1\linewidth]{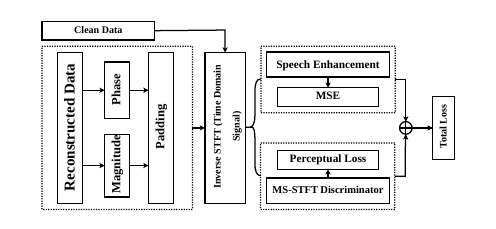}
    \caption{Proposed task aware speech enhancement using perceptual loss.}
    \label{fig:proposed}
\end{figure}


\subsection{Downstream Task: Speech Enhancement with Score-Based Generative Model}
We choose speech enhancement, that is, the task of extracting the clean speech signal from the background noise, as our downstream task of interest. This makes sense from a practical standpoint, considering that speech enhancement serves as a common initial block in multiple speech pipeline tasks in wireless AR/VR. To do speech enhancement, we use a score-based Langevin diffusion model \cite{welker22_interspeech} directly on the decoder output reconstructed STFT. The score output by the diffusion model learns to estimate the gradient of the log probability density of the noise-removed clean speech given the decoder output reconstructed noisy speech.

\subsubsection{Forward Process.}Here we refer to the decoder output STFT as the “noisy” spectrogram, and the clean speech STFT, produced by the downstream speech enhancement diffusion model, as the clean spectrogram. The forward process of the score-based Langevin diffusion model is implemented as an Ornstein-Uhlenbeck (OU) SDE:
\begin{equation}
    dx = -\theta (y - x) dt + \sigma(t) dw,
\end{equation}
where $x$ is the clean spectrogram, $y$ is the noisy spectrogram, $\theta$ is the stiffness parameter, $\sigma(t)$ is the time-dependent noise level, and $w$ is the Wiener process. It adds noise gradually to the spectrogram in subsequent timesteps as:
\begin{equation}
    \sigma(t) = \sigma_{\text{min}} \left( \frac{\sigma_{\text{max}}}{\sigma_{\text{min}}} \right)^t \sqrt{2 \log \left( \frac{\sigma_{\text{max}}}{\sigma_{\text{min}}} \right)}
\end{equation}
where $\sigma_{\text{min}}$ and $\sigma_{\text{max}}$ are the minimum and maximum noise levels.
\subsubsection{Score Network.} The score network is based on a speech enhancement diffusion model \cite{welker22speech}. It uses time embedding and conditioning the noisy spectrogram to output the score estimate (gradient of log probability). The score computation is given as: 
\begin{equation}
    s_{\theta}(x,y,t) = -\sigma(t)^2 \nabla_x \log p(x|y,t)
\end{equation}
where $s_{\theta}$ represents the score function, $\sigma^2(t)$ represents time-dependent noise schedule, $\nabla_x \log p(x|y,t)$ represents gradient of the log-probability of the clean spectrogram $x$ given the noisy spectrogram $y$. The score network learns to estimate $\nabla_x \log p(x|y,t)$. This is used to estimate the enhanced audio from the reconstructed noisy signal. The speech enhancement task loss component is computed based on denoising score matching as:
\begin{equation}
     L = \mathbb{E} \left[ \| s_{\theta}(x,y,t) + \sigma(t)^2 \nabla_x \log p(x|y,t) \|^2 \right]
\end{equation} 


\subsection{Perceptual Loss: Preserving Realism}
The last part of the loss function deals with preserving human-like speech, or realism, at the decoder output. Note that, to stay task-agnostic with the perceptual loss component, we intend to preserve realism at the decoder output, not at the downstream speech enhancement diffusion model output. The MS-STFT Discriminator \cite{défossez2022highfidelityneuralaudio}(Multi-Scale Short-Time Fourier Transform Discriminator) is a neural network architecture designed to extract perceptual features from audio data by leveraging multi-resolution frequency analysis and convolutional operations.

\subsubsection{Hierarchical Convolution Mapping.} To extract meaningful perceptual features from the decoder output speech STFT, we use a hierarchical convolution mapping\cite{NIPS2010_a0161022}. Hierarchical convolutional mapping involves applying a series of stacked convolutional layers to extract progressively abstract and meaningful features. Each convolutional operation extracts localized features from the spectrogram using a kernel as:
\begin{equation}
    y_{ij} = \sum_{m=0}^{k_h} \sum_{n=0}^{k_w} x_{(i+m)(j+n)} \cdot w_{mn},
\end{equation}
where \(x_{ij}\) is the input at position \((i, j)\), \(w_{mn}\) represents the kernel weights, and \(k_h, k_w\) denote the kernel height and width. This operation slides the kernel over the input, computing a weighted sum of overlapping regions to produce the output \(y_{ij}\). The use of progressively abstract feature extraction at the later layers of the hierarchical CNN is particularly suited for speech, which comprises hierarchical features, starting from localized phonemes, to broader semantic features across words or sentences spoken.

\subsubsection{Multi-Scale Discriminator}The MS-STFT Discriminator contains multiple discriminators, each configured with different STFT parameters. For each discriminator $D_i$:
\begin{equation}
    L_i, F_i = D_i(x),
\end{equation}
where \(D_i(x)\) represents the \(i\)-th discriminator applied to input \(x\), producing logits \(L_i\) and feature maps \(F_i\). This formulation captures both the decision output and intermediate perceptual features extracted by the discriminator. The input to the model is \( x \in \mathbb{R}^{B \times 1 \times T} \), producing complex spectrograms \( \mathbb{C}^{B \times 2 \times F \times T'} \). It uses \( N \)-layer 2D convolutions with kernel size \( k \), stride \( s \), dilation \( d \), and normalization to iteratively map features \( \mathbb{C}^{B \times C \times T' \times F} \to \mathbb{C}^{B \times C' \times T'' \times F'} \). Multiple discriminators at \( M \)-scale STFT resolutions capture hierarchical time-frequency features, yielding logits and feature maps across scales.

\subsubsection{Perceptual Loss.}We extract perceptual features of both the ground truth and the enhanced audio after passing it through the discriminator. The difference of their logits is then minimized to compute the perceptual loss. The logits are computed using MS-STFT Discriminator.
\(F_i\) at intermediate layers represent perceptual features of the audio. These features correspond to local energy distributions in the spectrogram, harmonic structures and transient details. For the \(l\)-th layer, the feature map \(F_l\) is defined as:
\begin{equation}
    F_l = \phi(W_l * F_{l-1} + b_l),  
\end{equation}
where \(\phi\) is the activation function, \(W_l, b_l\) are the weights and biases and \(*\) is the convolutional operator. These hierarchical features capture perceptual cues such as pitch, timbre, and temporal modulations.
\subsubsection{Logits as Output.}The final output logits \(L\) are computed after the last convolutional layer. Each discriminator produces a scalar logit \(L_i\) indicating the realism of the input audio:
\begin{equation}
    L = \text{Conv2D}(F_l).
\end{equation}
The logits are trained using adversarial objectives with real and generated signals as inputs:
\begin{equation}
    L_D = -\mathbb{E}[\log D(x_\text{real})] - \mathbb{E}[\log(1 - D(x_\text{fake}))].
\end{equation}
By employing multiple STFT configurations, the model captures audio features across a wide range of temporal and spectral resolutions. The convolutional layers learn hierarchical features, progressively refining perceptual representations and the normalization techniques ensure stable gradient flows, improving convergence. The overall model architecture has been illustrated in Fig.~\ref{fig:proposed}.


\section{Experiments}
\label{sec:experiments}
\begin{figure*}[t!]
    \centering
    \begin{subfigure}[t]{0.66\columnwidth}
        \centering
        \includegraphics[width=\textwidth]{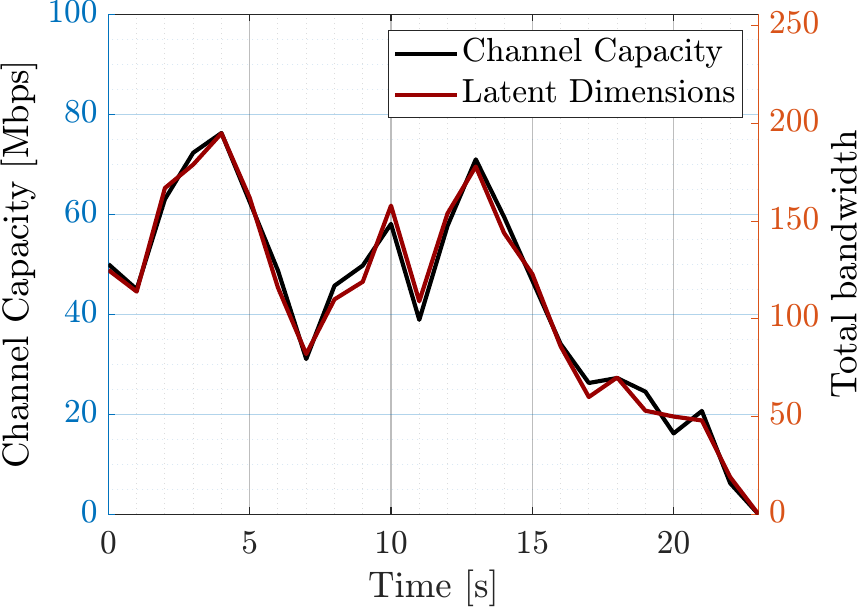}
        \caption{CSI aware bandwidth retention using NDPCA}
        \label{fig:bitrate}
    \end{subfigure}
    \hspace{-0.01\columnwidth}
    \begin{subfigure}[t]{0.66\columnwidth}
        \centering
        \includegraphics[width=\textwidth]{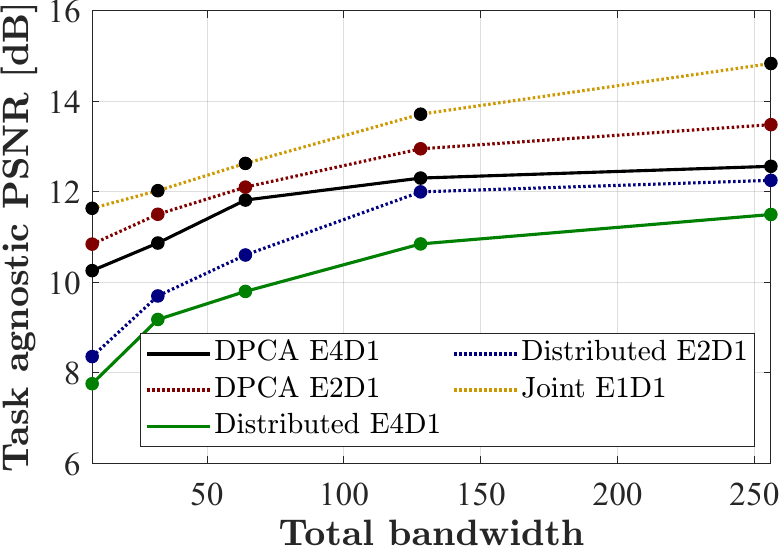}
        \caption{Total bandwidth vs. task agnostic PSNR for distributed autoencoder.}
        \label{fig:taskAgnostic}
    \end{subfigure}
    \hspace{-0.01\columnwidth}
    \begin{subfigure}[t]{0.66\columnwidth}
        \centering
        \includegraphics[width=\textwidth]{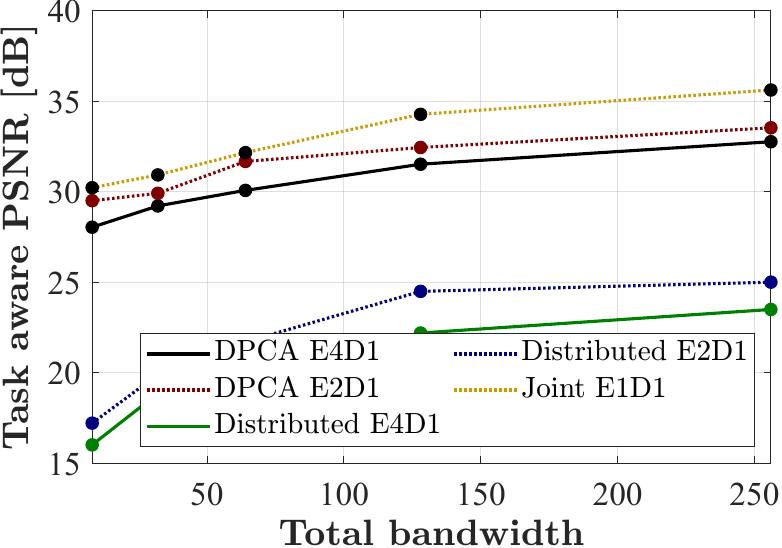}
        \caption{Total bandwidth vs. task aware PSNR for distributed autoencoder.}
        \label{fig:taskAware}
    \end{subfigure}

    \caption{\textbf{(a)} Efficient source coding using CSI-aware feedback vs. channel capacit, \textbf{(b)} Dimensions of latent space vs. task agnostic PSNR [db] for baseline E1D1, E2D1 and E4D1 autoencoders, \textbf{(c)} Dimensions of latent space vs. task aware PSNR [db] for baseline E1D1, E2D1 and E4D1 autoencoders.}
    \label{fig:example_grid}
\end{figure*}

\subsection{Baseline Methods}
The lack of prior work in task and perception-aware distributed source coding for speech data makes it impossible to make a fair comparison with the existing source coding baselines. Hence, we design the following two approaches as baselines.

\subsubsection{Joint Autoencoder (JAE)}
The joint autoencoder approach, which would be referred to as joint E1D1 henceforth, deals with the simple case where all the speech data from all the sources are combined as a single super source, and processed by a single encoder for source coding. This is the best case when it comes to dynamic bandwidth-based source coding rate adaptation, since all the information is available at a single source. Thus, in this case, we take the encoder output, followed by simple PCA, which chooses the top $k$ principal components, where $k$ is the maximum available bandwidth. This best case forms an upper bound on our proposed NDPCA-based distributed source coding architecture, where the individual sources cannot communicate with each other. For joint E1D1, the input spectral data includes magnitude and phase components and is represented as \(\mathbf{X} = [\mathbf{X}_m, \mathbf{X}_p] \in \mathbb{R}^{B \times 2 \times F \times T \times S}\), where \(B\) is the batch size, \(2\) represents the channels (magnitude and phase), \(F\) is the number of frequency bins, \(T\) is the number of time frames and $S$ is the number of sources. A single encoder processes the input \(\mathbf{X}\) to generate a latent representation \(\mathbf{Z}\), from which the top $k$ PCA components are transmitted over the channel, and decoded at the receiver. This process is mathematically expressed as:
\begin{equation}
    \mathbf{Z} = \text{Encoder}(\mathbf{X}),  \hat{\mathbf{Z}} = \text{PCA}(\mathbf{Z}),  \hat{\mathbf{X}} = \text{Decoder}(\hat{\mathbf{Z}}).
\end{equation}

\begin{figure}
    \centering
    \includegraphics[width=0.75\linewidth]{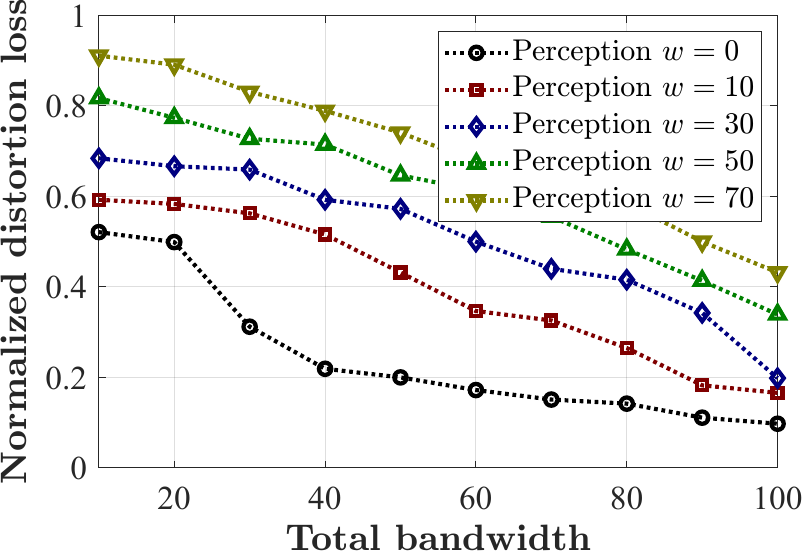}
    \caption{Experimental distortion versus total bandwidth under varying relative weights given to perception loss}
    \label{fig:rateDistortionPerception}
\end{figure}

\subsubsection{Distributed E2D1 \& E4D1 without NDPCA.}Our second category of baselines involves naive separated encoders for every individual source, or partially separate encoders combining a subset of the sources, without distributed PCA. Since in our experiments, we deal with speech recorded by 4 different mics (sources), we refer to these baselines as E4D1 and E2D1. E4D1 has a separate encoder for each of the mics, and each encoder processes the spectrogram of the recorded speech by that mic to extract latent representations. For the E4D1 architecture, the spectral data input consists of magnitude and phase components, represented as \(
\mathbf{X}_i = [\mathbf{X}_{i,m}, \mathbf{X}_{i,p}] \in \mathbb{R}^{B \times 2 \times F \times T}, \quad i \in \{1, 2, 3, 4\} \). Each input \(\mathbf{X}_i\) is processed by its corresponding encoder to produce latent representations, \(\mathbf{Z}_i\) as \(\mathbf{Z}_i = \text{Encoder}_i(\mathbf{X}_i), \quad i \in \{1, 2, 3, 4\}.\)The latent features are decomposed into private and shared components as \(\mathbf{Z}_i = [\mathbf{Z}_{i,p}, \mathbf{Z}_{i,s}], \quad i \in \{1, 2, 3, 4\}.\) and are then concatenated as:
\begin{equation}
    \mathbf{Z}_i = [\mathbf{Z}_{i,p}, \mathbf{Z}_{i,s}], \quad i \in \{1, 2, 3, 4\}.
\end{equation}
E2D1 follows the architecture along the same lines with a small difference of two encoders instead of four.

\subsection{Experimental Setup}
The experimental setup leverages a structured spectral dataset comprising clean and noisy audio signals. Clean spectral samples \( \mathbf{X}_{\text{clean}} \in \mathbb{R}^{1025 \times 600} \) were paired with their noisy counterparts \( \mathbf{X}_{\text{noisy}}^{(c)} \), sourced from four distinct channels, \( c \in \{3, 4, 5, 6\} \), ensuring consistent input dimensions through zero-padding where necessary. Training was performed with a batch size of \( N = 16 \) and a learning rate of \( \eta = 2 \times 10^{-4} \), over \( T = 100 \) epochs. Metrics such as mean squared error \( \mathcal{L}_{\text{MSE}} \), nuclear norm \( \| \mathbf{Z} \|_* \), cosine similarity, spectral SNR, magnitude loss \( \mathcal{L}_{\text{mag}} \), phase loss \( \mathcal{L}_{\text{phase}} \), and PSNR for both clean and noisy reconstructions were monitored. Training utilized PyTorch, executing on a CUDA-compatible GPU with a fixed random seed \( \text{seed} = 0 \) for reproducibility.

\subsection{Results}

\subsubsection{Task Agnostic: Comparison between Baselines and Proposed Algorithms.}
Fig.~\ref{fig:taskAgnostic} shows the PSNR achieved for the reconstructed speech versus the total bandwidth. In the task-agnostic setting, the total bandwidth is optimally allocated among the various sources involved using the proposed NDPCA-aided perception-aware source coding algorithm. We see from Fig.~\ref{fig:taskAgnostic} that the baseline Joint E1D1, which represents all source data being concatenated at one encoder, and thus maximum utilization of correlation of the different sources, achieves the upper bound PSNR values. Similarly, the baselines Distributed E4D1 and E2D1, where the 4 source mic data are all encoded individually, or assembled in pairs of 2 and encoded using 2 encoders, respectively, without any communication between the different encoders, and thus no NDPCA, forms the lower baseline on our proposed NDPCA-aided algorithm. As expected, the E2D1 cases perform better than the E4D1 cases, since pairs of sources are processed at one encoder for E2D1, creating higher correlation utilization than E4D1, where each source has its own encoder. The proposed algorithm with E2D1 achieves comparable performance to the upper bound Joint E1D1 (~91\%), thus demonstrating its efficiency and near-optimality. The proposed algorithms with distributed PCA E4D1 and E2D1 outperform the PSNR values obtained using the naive distributed source coding E4D1 and E2D1 by 18.8\% and 16.1\% respectively.

\subsubsection{Task Aware: Comparison between Baselines and Proposed Algorithms.}
Fig.~\ref{fig:taskAware} shows the results of the proposed algorithm, compared to the baselines, under the task and perception-aware setting. Once again, the trend is similar, with the proposed algorithms distributed PCA E4D1 and E2D1 outperforming the baselines E4D1 and E2D1 by 52.2\% and 46.9\% respectively, and achieving PSNR values comparable to the upper bound single encoder case (~98.1\%), especially under low bandwidth conditions. We see that the PSNR values obtained for the reconstructed speech spectrogram under the speech enhancement task and perception-aware setting significantly outperform the task-agnostic setting, the results for which are in Fig.~\ref{fig:taskAgnostic}. This clearly shows the effectiveness of our downstream task and perception aware loss function for the distributed speech source coding pipeline.

\subsubsection{Rate-distortion-perception analysis.}
Finally, we set out to experimentally analyze the effect of perception loss component on the 
Fig.~\ref{fig:rateDistortionPerception} shows the experimental rate-distortion levels obtained at various weights given to the perception loss component in the loss function, relative to the normalized speech enhancement task loss, which for our purposes becomes the task-aware distortion loss. As expected, the distortion obtained at a given total bandwidth, or rate, is higher for a higher relative weight given to the perception loss component, indicating that optimizing for distortion does not necessarily also optimize for higher realism and vice versa. Hence, it is upto use-case designers in future wireless AR/VR settings to asign relative importance levels to perception of decoded speech and accordingly tolerate the corresponding distortion level under a fixed total bandwidth.

\section{Conclusion}
\label{sec:conclusion}

In this paper, we introduce a novel task and perception-aware distributed speech source coding algorithm under dynamic bandwidth conditions. The NDPCA-aided autoencoder design, coupled with the direct downstream speech enhancement loss and perception loss-aware optimization, leads to superior PSNR values under a given bandwidth than those obtained for the following two baselines - (1)naive source coding where each of multiple sources is treated as an individual source and encoded without considering the correlation with the other sources, and (2) the task-agnostic case which is optimized for reconstruction loss only. 


\bibliography{main}

\newpage
\onecolumn
\appendix
\begin{center}
\LARGE{\textbf{Supplementary Materials}}
\end{center}
\section{\raggedright Dataset Preprocessing}
\vspace{0.5em}
For our system model, we used Chime 6 dataset~\cite{barker18_interspeech} which is a widely used for speech separation and recognition. The dataset consists of conversational speech audio with 8 speakers and 6 microphones spread across a room. The audio signals are aligned, compensated for frame drops and clock skew. Furthermore, all audio data are distributed as WAV files with a sampling rate of 16 kHz. Each session consists of the recordings made by the binaural microphones worn by each participant. We extract the noisy and clean audio signals for speakers and microphones through the given .json file for each start and end time for all the speaks as ground truth and the microphones as noisy data for that particular time. The extracted data is converted into .pkl with magnitude, phase, and parameters across time.
\begin{figure}[h!]
    \centering
    \includegraphics[width=0.5\columnwidth]{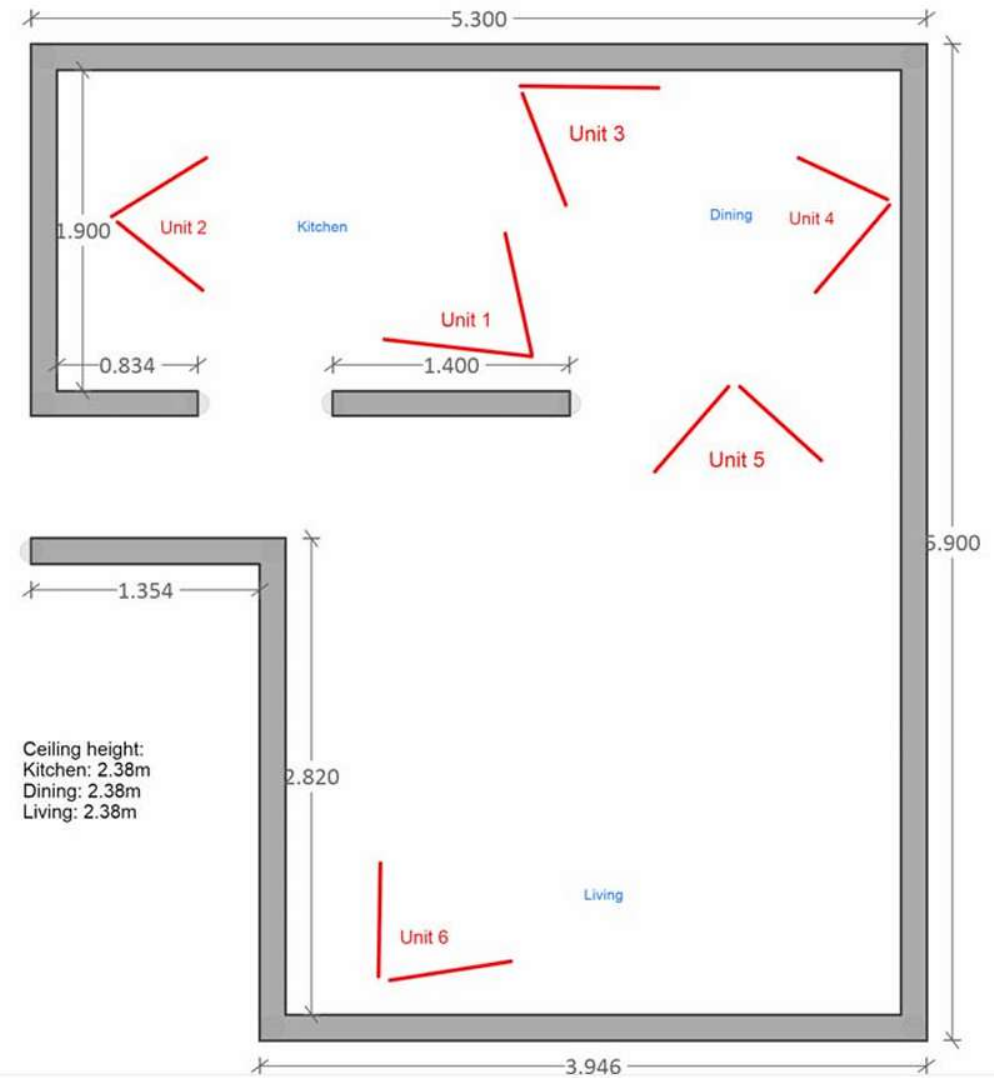}
    \caption{Floor plan for correlated audio data retrieval for distributed task-aware source coding~\cite{barker18_interspeech}.}
    \label{fig:floorplan}
\end{figure}
\section{\raggedright Dataset Retrieval Explanation}
\vspace{0.5em}
As shown in Fig.~\ref{fig:floorplan}, there are 8 speakers and 6 microphones in the room located at different positions as shown. Conversational audio during a party session is recorded through all the microphones. The social gatherings consist of close acquaintances who are encouraged to act casually and authentically. Multiple four-channel microphone systems capture and document these gatherings in their entirety. While participants can freely move between different areas, they must spend a minimum of half an hour in each location. The conversations are unscripted, allowing attendees to discuss any subject they choose, without following predetermined scenarios. For privacy protection, certain identifying information has been removed from the recordings during post-processing, in accordance with participant consent agreements.

The technical setup involves six Microsoft Kinect units positioned strategically throughout the space, ensuring that every area is monitored by at least two devices simultaneously. To ensure clear audio recording for transcription purposes, each participant is equipped with Soundman OKM II Classic Studio binaural microphones. These devices connect through a Soundman A3 adapter to personal Tascam DR-05 stereo recorders worn by the participants.

\section{\raggedright Spectrogram Distributions}
\vspace{0.5em}
To obtain spectrogram distributions of noisy and clean audio files, first we preprocess the data to manually separate the clean and noise wave file. we input a .wav file that contains raw waveform data and metadata describing speaker segments, including, start and end time along with session ID. Using timestamps from the JSON file, audio segments for a specific speaker are extracted by converting them into seconds. Using librosa, the waveform $y(t)$ the waveform is sampled between start time and end time with a sampling rate $f_s$ to yield:
\begin{equation}
    y[n] = y\left(\frac{n}{f_s}\right), \quad n = 0, 1, \ldots, N-1
\end{equation}
where N  is duration times $f_s$.
\subsubsection{Short-Time Fourier Transform.} 
\begin{figure*}[h!]
     \centering
     \begin{subfigure}[t]{0.4\columnwidth} 
         \centering
         \includegraphics[width=\linewidth]{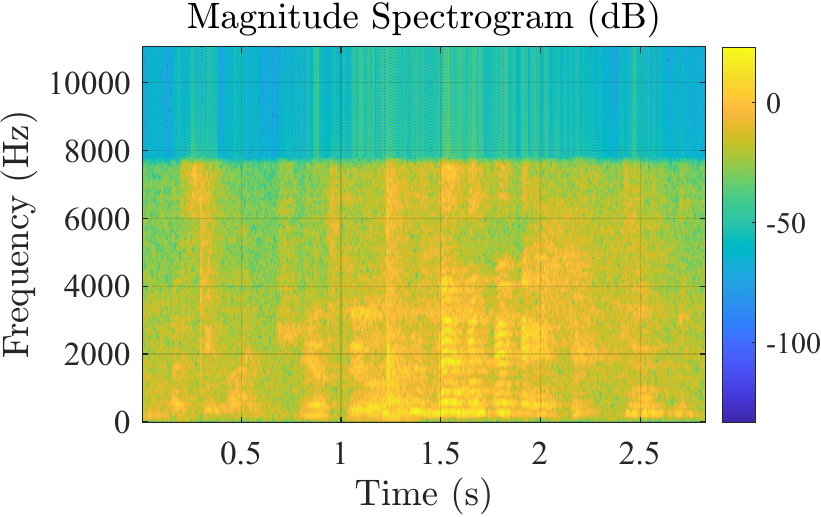}
         \caption{Clean magnitude spectrogram from speaker 8.}
         \label{fig:clean_magnitude}
     \end{subfigure}
     \hspace{0.02\columnwidth} 
     \begin{subfigure}[t]{0.4\columnwidth} 
         \centering
         \includegraphics[width=\linewidth]{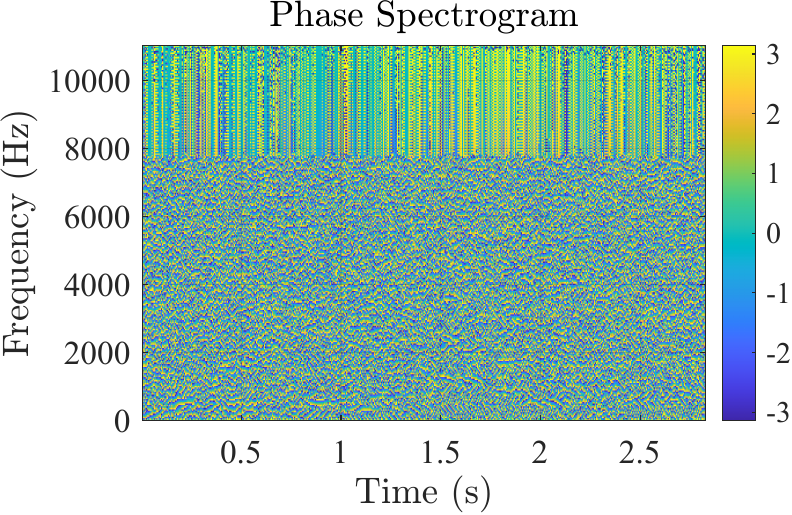}
         \caption{Clean phase spectrogram from speaker 8.}
         \label{fig:clean_phase}
     \end{subfigure}

     \vspace{0.02\columnwidth} 
     \begin{subfigure}[t]{0.4\columnwidth} 
         \centering
         \includegraphics[width=\linewidth]{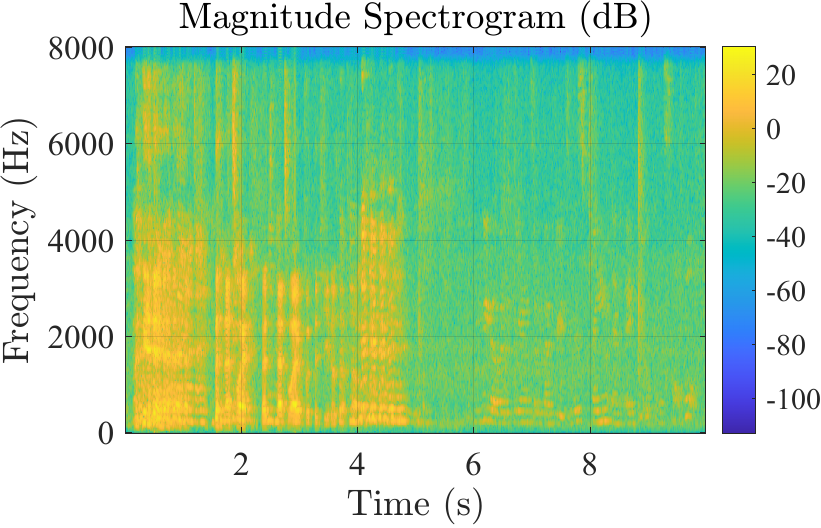}
         \caption{Noisy magnitude spectrogram from microphone 3.}
         \label{fig:noisy_magnitude}
     \end{subfigure}
     \hspace{0.02\columnwidth} 
     \begin{subfigure}[t]{0.4\columnwidth} 
         \centering
         \includegraphics[width=\linewidth]{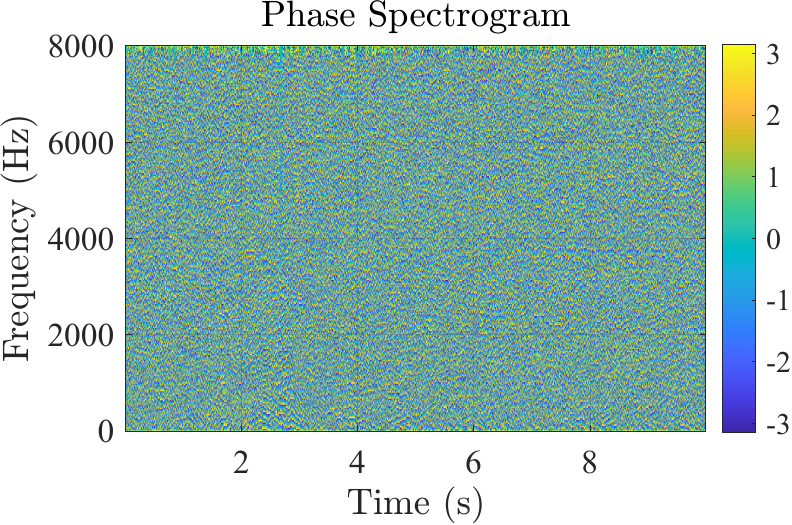}
         \caption{Noisy phase spectrogram from microphone 3.}
         \label{fig:noisy_phase}
     \end{subfigure}
     
     \caption{Magnitude and phase spectrogram distributions for clean and noisy audio signals with [FFT= 2048, Hop length= 512, and window length= 2048] parameters.}
     \label{fig:Spectrogram}
\end{figure*}

The core transformation involves converting the time domain signal $y[n]$ to time frequency representation \( D(f, t) \) to plot a spectrogram as:
\begin{equation}
    D(f, t) = \sum_{n=0}^{N-1} y[n] \cdot w[n - t \cdot H] \cdot e^{-j 2 \pi f n / N},
\end{equation}
where \( w[n] \) is the windowing function, \( H \) is the hop length, and \( t \) is the time frame index. As shown in Fig~\ref{fig:Spectrogram}, we obtain a magnitude and a phase response of the audio file, depicting and differentiating the characteristics of noisy and clean audio files. Fig.~\ref{fig:clean_magnitude} and Fig.~\ref{fig:clean_phase} represent the magnitude and phase of clean audio spectrograms with a clear time-frequency representation. The consistent patterns suggest the presence of tonal and structured features in the clean signal. The maximum magnitude of sound goes up to 0db for the clean signals. While the phase is typically harder to interpret directly, it remains consistent with the underlying clean signal. Fig.~\ref{fig:noisy_magnitude} and Fig.~\ref{fig:noisy_phase} demonstrate the noisy magnitude and phase response, where the noisy signal magnitude go as high as 20db. The frequency components are less distinct, and additional energy appears spread across frequencies, characteristic of added noise.

\subsubsection{3D Power Spectrgram.}
\begin{figure*}[t!]
     \centering
     \begin{subfigure}[t]{0.3\columnwidth} 
         \centering
         \includegraphics[width=\linewidth]{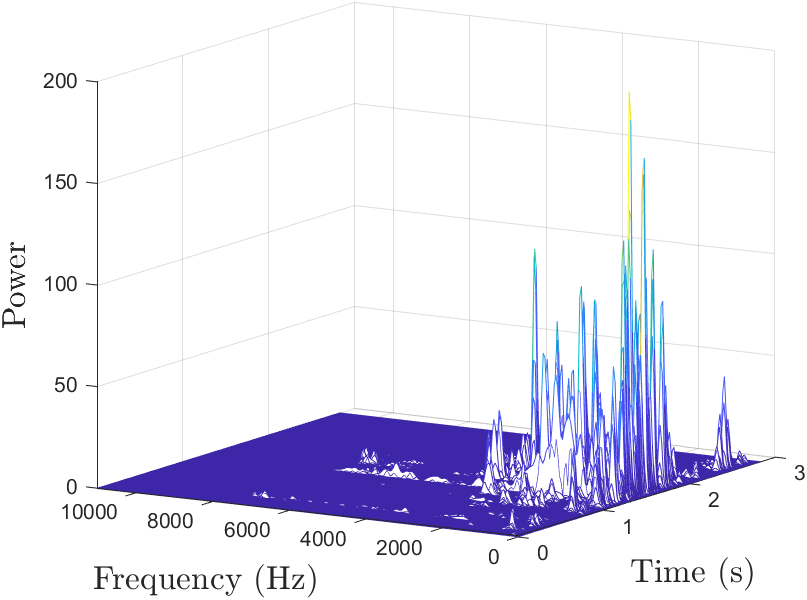}
         \caption{Power spectrogram for clean audio.}
         \label{fig:clean_3D}
     \end{subfigure}
     \hspace{0.02\columnwidth} 
     \begin{subfigure}[t]{0.3\columnwidth} 
         \centering
         \includegraphics[width=\linewidth]{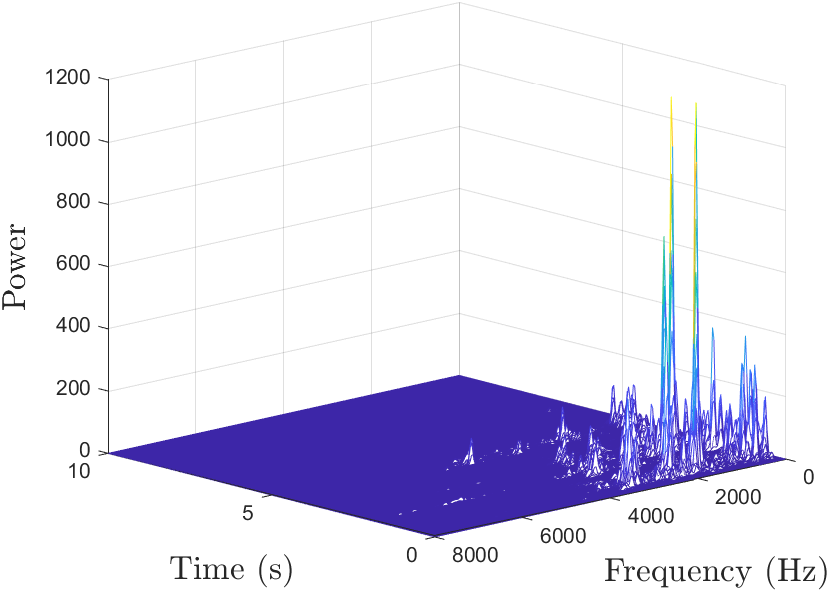}
         \caption{Power spectrogram for noisy audio.}
         \label{fig:noisy_3D}
     \end{subfigure}
     
     \caption{3D power spectrogram distribution of clean and audio signals against time [s] and frequency [Hz]}
     \label{fig:Spectrgrams_3D}
\end{figure*}

Fig.~\ref{fig:Spectrgrams_3D} shows the 3D power spectrogram distribution of the clean and the noisy signals. By applying Fourier transform and using mesh in Matlab we plot the power spectrogram of both noisy and clean audio signals as shown in Fig.~\ref{fig:clean_3D} and Fig.~\ref{fig:noisy_3D}. In Fig.~\ref{fig:clean_3D} we observe sharp and concentrated power peaks at specific frequencies, indicating well-defined tonal or harmonic components of the signal. Most of the power is concentrated in lower frequencies, which is typical for many natural sounds like speech. Fig.~\ref{fig:noisy_3D} shows a broader distribution of power across the frequency spectrum, indicating the presence of noise. Furthermore, the increased power levels, especially in high-frequency regions, suggests noise contamination.

\begin{table*}[h!] 
\centering
\caption{Comparison of task-agnostic and task-aware loss for various latent space dimensions}
\label{tab:my-table-combined}

\begin{tabular}{cc} 
    \begin{minipage}[t]{0.48\textwidth}
    \centering
    \caption{Task-agnostic loss for latent space dimension = 8}
    \begin{tabular}{>{\bfseries}lccc}
    \toprule
    \textbf{Model} & \textbf{Nuclear Loss} & \textbf{Cosine Loss} & \textbf{Spectral Loss} \\ \midrule
    E1D1           & 0.1672                & 0                    & 0.048                  \\ 
    E2D1           & 0.2650               & 0.297                & 0.08                  \\ 
    E4D1           & 0.4473                & 0.971                & 0.866                  \\ 
    \bottomrule
    \end{tabular}
    \end{minipage}
    &
    \begin{minipage}[t]{0.48\textwidth}
    \centering
    \caption{Task-agnostic loss for latent space dimension = 256}
    \begin{tabular}{>{\bfseries}lccc}
    \toprule
    \textbf{Model} & \textbf{Nuclear Loss} & \textbf{Cosine Loss} & \textbf{Spectral Loss} \\ \midrule
    E1D1           & 0.1351                & 0                    & 0.009                  \\ 
    E2D1           & 0.1650                & 0.0148               & 0.021                 \\ 
    E4D1           & 0.3173                & 0.767                & 0.317                  \\ 
    \bottomrule
    \end{tabular}
    \end{minipage} \\[1em]

    \begin{minipage}[t]{0.48\textwidth}
    \centering
    \caption{Task-aware loss for latent space dimension = 8}
    \begin{tabular}{>{\bfseries}lcc}
    \toprule
    \textbf{Model} & \textbf{Speech Enhancement Loss} & \textbf{Perceptual Loss} \\ \midrule
    E1D1           & 0.0000579                          & 0.3193                   \\ 
    E2D1           & 0.000145                          & 0.493                   \\ 
    E4D1           & 0.000163                          & 0.620                   \\ 
    \bottomrule
    \end{tabular}
    \end{minipage}
    &
    \begin{minipage}[t]{0.48\textwidth}
    \centering
    \caption{Task-aware loss for latent space dimension = 256}
    \begin{tabular}{>{\bfseries}lcc}
    \toprule
    \textbf{Model} & \textbf{Speech Enhancement Loss} & \textbf{Perceptual Loss} \\ \midrule
    E1D1           & 0.0000413                          & 0.3052                   \\ 
    E2D1           & 0.000081                          & 0.512                  \\ 
    E4D1           & 0.00034                          & 0.745                   \\ 
    \bottomrule
    \end{tabular}
    \end{minipage} 
\end{tabular}

\end{table*}
\begin{figure}[h!]
    \centering
    \includegraphics[width=0.5\linewidth]{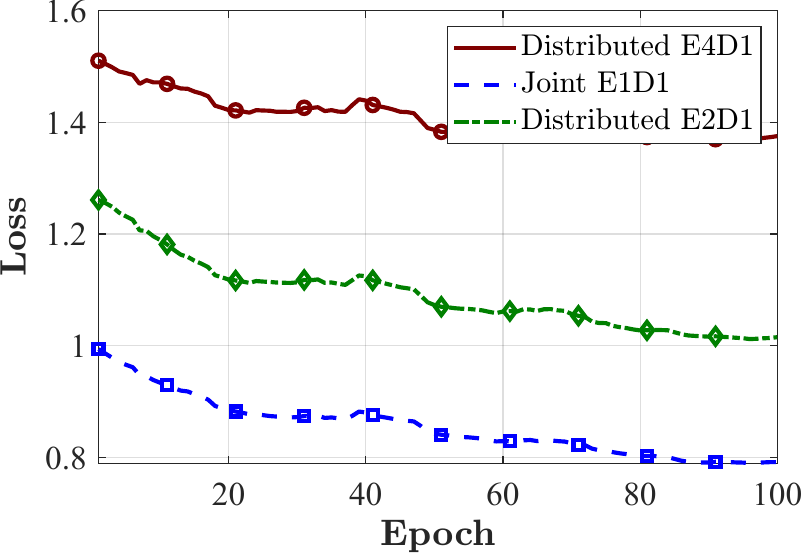}
    \caption{Task agnostic loss for different baseline models.}
    \label{fig:task_agnostic}
\end{figure}
\section{\raggedright Training Hierarchy and Loss values}
In this section, we observe the convergence and the loss for both task agnostic and task aware perceptual pipelines.
\subsubsection{Loss Convergence.}From Fig.~\ref{fig:task_agnostic}, we observe that E2D1 performs better than E4D1 on the overall loss curve and converges to a lower value. As expected, the E2D1 cases perform better than the E4D1 cases since pairs of sources are processed at one encoder for E2D1, creating higher correlation utilization than E4D1, where each source has its own encoder. Mathematically, consider the latent representations \( \mathbf{z}_i \in \mathbb{R}^{d_i} \) produced by the encoders. In E4D1, the four encoders independently encode features as \( \mathbf{z}_1, \mathbf{z}_2, \mathbf{z}_3, \mathbf{z}_4 \), allowing the system to represent input modalities or perspectives as a concatenated vector \( \mathbf{z} = [\mathbf{z}_1; \mathbf{z}_2; \mathbf{z}_3; \mathbf{z}_4] \in \mathbb{R}^{d} \), where \( d = d_1 + d_2 + d_3 + d_4 \). However, this independent encoding limits the correlation utilization between sources, resulting in a suboptimal alignment of features. In E2D1, the encoders process pairs of sources, producing latent representations \( \mathbf{z}_{1,2}, \mathbf{z}_{3,4} \in \mathbb{R}^{d_1 + d_2} \). This structure enables better correlation utilization between paired sources, minimizing the projection error \( \epsilon = \| \mathbf{x} - P_\mathcal{Z}(\mathbf{x}) \| \), where \( P_\mathcal{Z}(\mathbf{x}) \) is the projection onto the latent space. The improved inter-source alignment in E2D1 results in richer feature extraction and higher reconstruction accuracy. Furthermore, the proposed algorithm with E2D1 achieves comparable performance to the upper bound Joint E1D1, demonstrating its efficiency and near-optimality. By leveraging pairwise correlation, E2D1 achieves enhanced representational capacity, significantly reducing latent divergence \( \mathbb{E}[\| \mathbf{z}_i - \mathbf{z}_j \|] \). This improves both cross-reconstruction and global decoding consistency, thus outperforming E4D1 in tasks requiring efficient utilization of correlated features.  \\
\subsubsection{Loss values.}The loss values are shown in Table. 1-4, for task aware and task agnostic loss. The nuclear loss corresponds to reconstruction loss and cosine loss is for correlation. The correlation loss is 0 for joint autoencoders, because they can communicate with each other and are concatenated before encoding. The tables justify the above model performance explanations.

\end{document}